\documentclass[twocolumn,aps,prd,preprintnumbers]{revtex4}

\usepackage{graphicx}
\usepackage{amsmath}
\usepackage{amssymb}
\usepackage{hyperref}
\begin{document}
\preprint{DO-TH 05/11}
\title{Baryon and lepton numbers in two scenarios of leptogenesis.}
\author{A.Kartavtsev}
\email{akartavt@het.physik.uni-dortmund.de}
\affiliation{Institut f\"{u}r Physik, Universit\"{a}t Dortmund, D-44221 
Dortmund, Germany}

\begin{abstract}
Baryon and lepton numbers of the Universe in  leptogenesis with Dirac 
neutrino and leptogenesis with Majorana neutrino scenarios are considered. 
It is shown that despite quite different features of Dirac and Majorana fermions 
both scenarios yield the same relation among the initial lepton and the final 
baryon asymmetries. Moreover right--handed neutrinos in the leptogenesis with 
Dirac neutrino scenario have very little impact on the effective number 
of relativistic degrees of freedom, constrained by BBN. Thus the two scenarios
are similar from the cosmological point of view. It is also pointed out that in 
thermal equilibrium the $3B+L$ sum is zero for left--handed fermions.
\end{abstract}

\maketitle

\section{\label{introduction}Introduction.}
The observed baryon asymmetry of the Universe is one of the most
interesting problems of particle physics and cosmology. The cosmological 
baryon excess can be generated in decays of heavy particles, provided 
that the three Sakharov conditions \cite{Sakharov:1967dj} are satisfied: baryon 
(or baryon minus lepton) number is not conserved;  C and CP are violated; 
the decay processes are out of equilibrium.

It has been proven to be difficult to generate excess of baryons through 
direct violation of baryon number. Discovery of anomalous electroweak 
processes \cite{'tHooft:1976up}, violating baryon ($B$) and lepton ($L$) 
numbers but conserving the difference $B-L$, led to the widely adopted
scenario of leptogenesis; the initial lepton number asymmetry, created at
a GUT scale, is later converted into baryon asymmetry by  anomalous 
processes (refereed to as sphalerons in what follows).
The remaining lepton number is distributed among charged leptons and 
neutrinos.  

According to the scenario suggested by M. Fukugita  and T.  Yanagida
\cite {Fukugita:1986hr} lepton number asymmetry is generated in 
the lepton  number violating and CP violating decay of heavy Majorana neutrino.  
After the electroweak (EW) phase transition 
conventional neutrinos acquire naturally small Majorana masses due to the 
see--saw mechanism. This scenario is refereed to as leptogenesis with Majorana 
neutrinos.

In an alternative scenario suggested in \cite{Dick:1999je} total lepton
number is conserved, but CP violating decay of heavy particle  results 
in nonzero lepton number for left--handed particles, and equal in magnitude 
but opposite in sign lepton number for right--handed particles. The negative 
lepton number stored in the left--handed neutrinos is converted by the 
sphalerons into positive baryon number. This scenario is refereed to  as 
leptogenesis with Dirac neutrinos.

These two scenarios are compared here. In particular, the observed baryon 
asymmetry of the Universe and effects of additional relativistic degrees of 
freedom are discussed.

The sphaleron process is a crucial ingredient of both scenarios.
It is pointed out that if the sphalerons are in thermal equilibrium, then 
the  $3B+L$ sum is zero for left--handed fermions.

Majorana and Dirac neutrinos have quite different properties. The  Dirac 
neutrino is a four--component fermion and a definite lepton number can be 
assigned to its left-- and right--handed components. The Majorana neutrino is 
a two--component fermion and the Majorana mass term violates lepton number by 
two units. Equivalently, the Majorana neutrino does not carry any conserved 
charges and does not have a chemical potential. 

The relation between initial $B-L$ asymmetry generated at the GUT scale and the 
observed baryon asymmetry of the Universe was studied in 
\cite{Harvey:1990qw, Dreiner:1992vm} using relations among the particles 
chemical potentials \footnote{An analysis applicable also in the $T\gg\phi$ 
limit, where $\phi$ is the temperature dependent Higgs expectation value, has 
been performed in \cite{Laine:1999wv}}.  If the statement about the Majorana 
neutrino zero chemical potential was applicable in the 
early Universe, then the initial $B-L$ asymmetry would be completely washed
out below the electroweak phase transition. However, the rate of lepton number 
violating processes, estimated here, turns out to be  much smaller than 
expansion rate of the Universe and thus the neutrino can be assigned an 
effective chemical potential. 

In the scenario of leptogenesis with Dirac neutrino  conserved total
lepton number is  redistributed among the left-- and the right--handed components
in the processes with Higgs and gauge boson exchange. Due to smallness of the neutrino 
mass these processes are too slow to bring lepton number of the left-- and 
right--handed neutrinos to equilibrium.

Thus both scenarios yield the same relation between the initial lepton 
and the final baryon asymmetry. Moreover, since the right--handed neutrino in
the scenario of leptogenesis with Dirac neutrino carries only a small fraction 
of energy density of the Universe, it has very little impact on the effective 
number of relativistic degrees of freedom, constrained by BBN, and therefore 
the two scenarios are similar from the cosmological point of view.

\section{\label{processes}Electroweak processes.}
Assuming a thermal distribution, the number density of fermions (bosons) of 
mass $m$ is given by the Fermi--Dirac (Bose--Einstein) distribution
\begin{equation}
\label{densitydistr}
n_\pm=\int\frac{{\rm d} {\bf p}}{(2\pi)^3}\frac{g}{\exp[(E_{\bf p}\mp\mu)/T]\pm 1},
\end{equation}
where $\mu$ is the particle chemical potential and 
$g$ is the number of internal degrees of freedom ($g=1$ for massless Weyl 
fermions and $g=2$ for massless vector bosons). 
Assuming that $\mu/T$
is small and expanding (\ref{densitydistr}) in powers of the particle mass $m$  
we find for excess of particles over antiparticles
\begin{align}
\label{deltaN}
n_{+}-n_{-}&=\frac{gT^3}{6}\frac{\mu}{T}\, c_f\left( \frac{m}{T}\right) 
({\rm fermions}),\\
n_{+}-n_{-}&=\frac{gT^3}{3}\frac{\mu}{T}\, c_b\left( \frac{m}{T}\right)
({\rm bosons}).
\end{align}
where the functions $c_f$ and $c_b$ are given by
\begin{align*} 
c_f\left( \frac{m}{T}\right) \approx1-\frac{3}{2\pi^2}\frac{m^2}{T^2},\,
c_b\left( \frac{m}{T}\right)\approx 1-\frac{9}{2\pi^2}\frac{m}{T}
+\frac{3}{4\pi^2}\frac{m^2}{T^2}
\end{align*}
Baryon asymmetry of the Universe, which is usually expressed as 
ratio of the baryon density to the entropy density of the Universe, 
is given by
\begin{equation}
Y_B=\frac{n_B-n_{\bar{B}}}{s}=(0.6-1)\cdot 10^{-10}
\end{equation}
In the radiation dominated Universe the entropy density  $s=2\pi^2g_{*}T^3/45$, 
where $g_*$ counts the total number of relativistic degrees of freedom, and 
therefore $\mu/T\sim Y_B$ is sufficiently small to justify the use of eq. 
(\ref{deltaN}).

As the sphaleron processes switch off immediately 
or slightly below the electroweak phase transition,
only processes which are in equilibrium in the 
vicinity of  $T_C\sim 300$ GeV are important for 
baryon and lepton number formation.
Success of the standard model (SM) in explaining 
low--energy phenomena suggests, that all particles
besides the known ones are too heavy to be in thermal 
equilibrium at temperatures close to the critical. 

The standard model consists of $N$ generations of quarks and leptons, fields of 
$SU_C(3)\otimes SU_L(2)\otimes U_Y(1)$ gauge group and $m$ Higgs 
doublets \footnote{For instance, in extensions of the standard model  
based on the $E_6$ gauge  group each generation contains  two Higgs doublets.
Supersymmetric extension of the SM necessarily contain
at least two Higgs doublets responsible for masses
of the up and the down quarks. However, if quarks of given charge 
are  coupled to more than one light Higgs, strong flavor--changing 
neutral currents arise. Numerical values are given here only 
for $m=1$.}.	
Above the electroweak  phase transition  vacuum 
expectation value of the Higgs boson is zero and all the 
known fermions are massless. As temperature drops 
below the critical one, the fermions acquire masses. 
In the vicinity of $T_C$ the mass to temperature ratio is 
very small for all fermions but the $t$--quark. 

Since the $SU_C(3)$ symmetry is exact 
at any temperature, chemical potentials of the 
components of color triplets are equal,
i.e. chemical potential of the gluon fields is zero. 

Above the electroweak phase transition the $SU_L(2)$
symmetry is unbroken and components of   weak 
doublets have the same chemical potentials. Therefore
chemical potentials of the  $W$ bosons are zero. 
Chemical potential of the  $B^0$ gauge boson is zero
because it is neutral.  

Below the phase transition chemical potentials of the components 
of weak doublets are no longer equal and the $W^{-}$  boson 
acquires a nonzero chemical potential denoted by $\mu_W$. 
Since the photon and the $Z$ boson are neutral, they do not 
have  chemical potentials.

Rapid flavor--changing interactions in the quark sector 
assure that chemical potentials of quarks of given 
charge and chirality are the same for all flavors and colors both above and 
below the phase transition,  so that  only  four chemical 
potentials should be introduced: $\mu_{uL}$ and $\mu_{dL}$ for 
left--handed and  $\mu_{uR}$ and $\mu_{dR}$ 
for right--handed quarks of a given color 

In the absence of rapid flavor--mixing interactions chemical potentials 
$\mu_{iL}$ and $\mu_{iR}$ of charged leptons and chemical 
potentials $\mu_i$ of left--handed neutrinos are in general different. 
The phenomenon of neutrino oscillation, confirmed experimentally,
suggests that below the phase transition flavor--mixing 
interactions in the neutrino sector may be sufficiently fast to 
assure $\mu_{iL}=\mu_L$, $\mu_{iR}=\mu_R$ and $\mu_i=\mu_\nu$
equalities. 
Above the phase transition this is, however, not the case. 
The reason is the following. The 
source of the light neutrino mass is  $\lambda \bar{\nu}_R\nu_L H$ term.  
In the case 
of leptogenesis with Majorana neutrino  mass of the right--handed neutrino 
$\nu_R$ is many orders of magnitude bigger than $T_C$, and consequently 
flavor--changing interactions involving this vertex 
are out of equilibrium. In the case of leptogenesis 
with Dirac neutrino the coupling $\lambda$ is very small (as is required
by the small neutrino mass) and flavor--changing interactions
involving the Higgs exchange  are out of equilibrium as well. 

Finally, it is assumed, that mixing between the Higgs 
doublets assures the equality of their
chemical potentials: $\mu_{-}$ for all charged scalars and 
$\mu_0$ for the neutral ones.
As the temperature drops below $T_C$, the neutral Higgs develops nonzero
vacuum expectation value and breaks the $SU_L(2)$ symmetry. Since 
the neutral physical Higgs does not carry any conserved quantum numbers
its chemical potential $\mu_0$ is zero below the electroweak phase transition. 

The electroweak interactions are in thermal equilibrium down to about 
$T_{dec}\simeq 2$ MeV, which imply some useful relations among the chemical 
potentials \cite{Harvey:1990qw} partially discussed above.
\begin{subequations}
\label{reactions}
\begin{eqnarray}
\mu_W&=&\mu_{-}+\mu_{0}\quad (W^-\leftrightarrow H^-+H^0)\\
\mu_{dL}&=&\mu_{uL}+\mu_W\quad (W^-\leftrightarrow \bar{u}_L+d_L)\\
\mu_{iL}&=&\mu_i+\mu_W\quad (W^-\leftrightarrow \bar{\nu}_{iL}+e_{iL})\\
\mu_{uR}&=&\mu_{uL}+\mu_0\quad (H^0\leftrightarrow \bar{u}_L+u_R)\\
\mu_{dR}&=&\mu_{uL}+\mu_W-\mu_0\quad (H^0\leftrightarrow d_L+\bar{d}_R)\\
\mu_{iR}&=&\mu_i+\mu_W-\mu_0\quad (H^0\leftrightarrow e_{iL}+\bar{e}_{iR})
\end{eqnarray}
\end{subequations}
Using the relations  above, one can express baryon and lepton  number, 
as well as electric charges, in terms of just $3+N$ chemical potentials. 
Furthermore it is convenient to introduce the following notation:
\begin{align*}
c_{-}&=c_b(m_{\phi_{-}}/T)\\
c_{W}&=c_b(m_W/T)\\
c_{\ell_i,u_i,d_i}&=c_f(m_{\ell_i,u_i,d_i}/T).
\end{align*}
The Higgs bosons are massive both above and below the phase transition, 
and therefore  $c_{-}$  differs from unity at any temperature \footnote{In the 
case of the standard model charged components of the Higgs doublet are eaten 
up by the $W$ and $c_{-}$ is identically zero below the phase transition.}.
Fermions and gauge bosons  are massless above the phase transition, 
and consequently $c_W$ and $c_{\ell_i}$, $c_{u_i}$, $c_{d_i}$ are equal 
to unity at $T>T_C$. Deviation from unity at $T<T_C$ is small and is 
parametrized by
\begin{align*}
N\Delta_{\ell,u,d}&=N-\sum_ic_{\ell_i,u_i,d_i},\\
\Delta_\mu &=\mu-\sum_ic_{\ell_i}\mu_i,
\end{align*}
where $\mu\equiv\sum_i\mu_i$ and $N$ is the number of generations. 
In the presence of fast flavor--changing 
interactions in the lepton sector  $\Delta_\mu=\mu\Delta_\ell$.

Chemical potentials $\mu_i$ of the neutrinos appear only in  the combinations 
$\mu$ and $\Delta_\mu$ and we can express total baryon and lepton 
numbers  in terms of just five chemical potentials. Omitting an overall 
coefficient we find
\begin{subequations}
\label{BandLgeneral}
\begin{align}
\label{Bgeneral}
B&=N(1-\Delta_u)(\mu_{uL}+\mu_{uR})+N(1-\Delta_d)(\mu_{dL}+\mu_{dR})\nonumber\\
&=2N(2-\Delta_u-\Delta_d)\mu_{uL}+2N(1-\Delta_d)\mu_W\\
\label{Lgeneral}
L&=\sum_i[\mu_i+c_{\ell_i}(\mu_{iL}+\mu_{iR})]=3\mu-2\Delta_\mu\nonumber\\
&+2N(1-\Delta_\ell)\mu_W-N(1-\Delta_\ell)\mu_0
\end{align}
\end{subequations}
To keep  track of  the number of different species in the plasma it 
is useful to introduce electric charges of up and down quarks and 
charged leptons
\begin{subequations}
\label{charges}
\begin{align}
Q_u&=2N(1-\Delta_u)(2\mu_{uL}+\mu_0)\\
Q_d&=-N(1-\Delta_d)(2\mu_{uL}+2\mu_W-\mu_0)\\
Q_l&=2\Delta_\mu-2\mu-2N(1-\Delta_\ell)\mu_W+N(1-\Delta_\ell)\mu_0
\end{align}
\end{subequations}
In the massless limit \cite{Harvey:1990qw} the combined charge of the Higgs 
and the gauge bosons 
\begin{equation*}
Q_{W+H}=-2(m+2)\mu_W+2m\mu_0
\end{equation*}
is given by one and the same formula both above and below the phase transition. 
When mass effects are taken into account  $c_{-}$ and $c_W$ are not 
equal and such a representation is no longer  possible.

Above the EW phase transition $\mu_W=0$ and  we obtain
\begin{equation}
\label{Qbosonabove}
Q_{H+W}=2mc_{-}\mu_0
\end{equation}
Below the phase transition two electrically charged 
components of the Higgs fields are ``eaten up'' by the $W$. 
As  massive vector boson has three spin degrees of freedom,
and $\mu_0$ is zero at this stage, the expression for the combined 
charge takes the form
\begin{equation}
\label{Qbosonbelow}
Q_{H+W}=-2( (m-1)c_{-}+3c_W) \mu_W
\end{equation}
In the case of the standard model ($m=1$) the physical Higgs is neutral
and only the  $W$--boson contributes to the charge density.

\section{\label{sphalerons}Sphaleron processes.}
An important ingredient of leptogenesis is the electroweak anomaly 
which  violates $B$ and $L$ but conserves $B-L$ quantum number. 
Let us first consider the case of only left--handed fermions above the 
electroweak phase transition. As the chemical potentials of  up and down 
components of electroweak doublets are equal and fermion masses 
are zero, the free energy ${\cal F}$  is given by
\begin{equation*}
{\cal F}=2\sum_iF(\mu_i)+6NF(\mu_{uL})\propto 2T^2\sum_i\mu_i^2+6 T^2 N\mu_{uL}^2
\end{equation*}
Baryon and lepton number densities are obtained by differentiating the free 
energy with respect to $\mu_i$ and $\mu_{uL}$
\begin{equation*}
L_L=\sum_i\frac{\partial{\cal F}}{\partial \mu_i}\propto4T^2\mu,\quad
B_L=\frac13\frac{\partial {\cal F}}{\partial \mu_{uL}}\propto4T^2N\mu_{uL}.
\end{equation*}
From the fact that  the sphalerons  conserve baryon minus lepton quantum
number separately for each generation \cite{Rubakov:1996vz} it follows that 
$d\mu_i=d\mu_{uL}$. Minimization of the free energy yields
\begin{equation}
\label{Lplus3B}
\frac{d {\cal F}}{d\mu_{uL}}\propto 4T^2(3N\mu_{uL}+\mu)\propto 3B_L+L_L=0
\end{equation}
In other words, if sphaleron processes are in equilibrium  sum
of the   lepton and thrice the baryon number stored in the 
left--handed fermions is zero.
\begin{figure}[h]
\includegraphics[width=7cm]{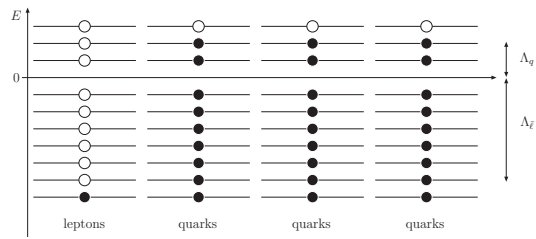}
\caption{\label{levels}Illustration to conversion of lepton number into 
baryon number by the electroweak anomaly. $\Lambda_{\bar{\ell}}$ and $\Lambda_q$ 
are (positive) maximal energies of antileptons and quarks.}
\end{figure}

To illustrate this conclusion let us assume that initial 
baryon number is zero and total lepton number is negative, 
as it is the case for leptogenesis. 
Number and energy densities of leptons and quarks are given by
\begin{equation*}
n_{\bar{\ell}}\propto \Lambda_{\bar{\ell}}^3,\quad n_{q}\propto3\Lambda_q^3,\quad
E_{\bar{\ell}}\propto \Lambda_{\bar{\ell}}^4,\quad E_{q}\propto 3\Lambda^4_q
\end{equation*}
In the course of sphaleron transitions empty negative lepton levels and filled
negative quark levels cross the zero from below so that number of antileptons 
is decreased and number of baryons is increased (Fig. \ref{levels}).
While both  $\Lambda_{\bar{\ell}}$ and $\Lambda_q$ change, the sum 
$\Lambda_{\bar{\ell}}+\Lambda_q$ remains constant. 
\begin{equation*}
\frac{d E}{d \Lambda_q}\propto 3\Lambda_q^3-\Lambda_{\bar{\ell}}^3\propto n_q- 
n_{\bar{\ell}}\propto 3B_L+L_L=0
\end{equation*}
The energy density of the system reaches its minimum when total number of 
antileptons is equal to \emph{total} number of quarks,
i.e. the baryon number is minus one third of the lepton number.

In a more general case it is convenient to use the fact that 
for equilibrium reactions sum of chemical potentials of the incoming 
particles is equal to that of outgoing. The sphaleron processes correspond
to the creation of $(u_{iL}d_{iL}e_{iL}\nu_{iL})$ states of each generation 
\cite{Rubakov:1996vz} out of the vacuum  \footnote{The created object is a 
$SU_L(2)$ singlet because both up-- and down components of weak doublets are 
created out of the vacuum and a color--singlet because quarks of all the three 
colors are created simultaneously.}. Therefore, as long as the sphaleron 
processes are in thermal equilibrium, the following relation among the chemical 
potentials is enforced:
\begin{equation}
\label{sphalerons1}
3N(\mu_{uL}+\mu_{dL})+\sum_i(\mu_{iL}+\mu_i)=0,
\end{equation}
where the factor of three is due to the three color degrees of freedom,
while summation over generations takes into account that fermions
of all the generations are created simultaneously.
Relations (\ref{reactions}) allow to  rewrite (\ref{sphalerons1}) as follows:
\begin{equation}
\label{sphalerons2}
3N\mu_{uL}+2N\mu_W+\mu=0.
\end{equation}
Above the electroweak phase transition (i.e. in the limit of $\mu_W=0$)
the last equality  coincides  with the eq. (\ref{Lplus3B}).

\section{\label{above}Above the phase transition.}
Above the phase transition the left--handed neutrino is a 
massless Weyl fermion in both scenarios.
Since the gauge bosons and the fermions are
massless, the anomalous electroweak processes are not 
suppressed. As the chemical potential of the $W$ is zero and $\Delta_\mu=0$
baryon and lepton numbers are proportional to the \emph{total} 
$B-L$ asymmetry. Requiring that  total electric
charge $Q$ is zero and making use of the relation implied 
by sphaleron transitions we find \cite{Harvey:1990qw, Dreiner:1992vm}
\begin{subequations}
\begin{align}
\label{Babove}
B&=\hphantom{-}\frac{8N+4mc_{-}}{22N+13mc_{-}}(B-L)\approx\hphantom{-} 0.36\, 
(B-L),\\
\label{Labove}
L&=-\frac{14N+9mc_{-}}{22N+13mc_{-}}(B-L)\approx-0.64\, (B-L),
\end{align}
\end{subequations}
where for the numerical  evaluation we shall choose $N=3$ and assume only one 
Higgs doublet  with  the Higgs mass of 400 GeV at $T_C$. While the $3B+L$ 
number stored in left--handed particles is zero, neither total $3B+L$ nor 
total $B+L$ are zero.

Electric charges stored in different species  are given by
\begin{subequations}
\begin{align}
Q_u&=\hphantom{-}\frac{4mc_{-}}{22N+13mc_{-}} (B-L)\approx\hphantom{-} 0.03\,(B-L)\\
Q_d&=-\frac{8N+2mc_{-}}{22N+13mc_{-}} (B-L)\approx-0.34\,(B-L)\\
Q_l&=\hphantom{-}\frac{8N+6mc_{-}}{22N+13mc_{-}} (B-L)\approx\hphantom{-} 0.37\,(B-L)
\end{align}
\end{subequations}
i.e., above the electroweak phase transition there is an excess of down quarks
and total electric charge of  the leptons is positive.

\section{\label{dirac}Leptogenesis with Dirac neutrinos.}
According to the scenario suggested in \cite{Dick:1999je}, even in a model
that conserves lepton number, a  CP violating decay of heavy particle can
result in a nonzero lepton number for left--handed particles, and  
equal in magnitude but opposite in sign lepton number for right--handed 
particles. The negative lepton number stored in the 
left--handed fermions is  later converted into positive baryon number by the 
sphalerons .

In this scenario below the phase transition the neutrino is a four--component
Dirac spinors with very small mass. Coupling of its right--handed component 
to the neutral Higgs is given by 
\begin{equation}
\label{diraclagr}
{\cal L}=\frac{g}{2}\frac{m_\nu}{M_W}\bar{\nu}\nu H_0
\end{equation}
In (\ref{diraclagr}) the Yukawa coupling is expressed as a ratio  
of the neutrino mass to the Higgs VEV. The later, in turn, is proportional 
to the $W$ mass. 

\begin{figure}[h]
\begin{center}
\includegraphics{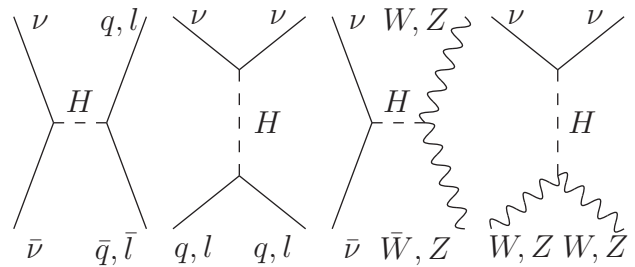}
\end{center}
\caption{\label{diracwashout}Higgs--mediated processes which equilibrate lepton 
asymmetry stored in the left-- and right--handed neutrinos.}
\end{figure}

Negative and positive lepton numbers stored,  respectively,  in the left--  and 
right--handed neutrinos can partially equilibrate. However, as is clear from 
(\ref{diraclagr}), rate of the equilibration processes is strongly suppressed 
by the neutrino to the $W$ mass ratio.  To be more quantitative, let us evaluate 
the  corresponding diagrams. In the  processes depicted in Fig. 
\ref{diracwashout} (and their charge conjugate) total lepton number
is conserved, but lepton asymmetry is redistributed among the left-- and 
right--handed neutrinos.  

The leading contribution at $T\sim T_C$ is due to the  top--quark, whose Yukawa 
coupling is of order of unity, and due to the $W$ and $Z$ bosons.
Reduced cross sections of the  diagrams with the top--quark are given by
\begin{align}
\label{dirac_s}
\hat{\sigma}_s^{t}&=\frac{g^4}{32\pi}
\left(\frac{m_\nu}{M_W}\frac{m_{t}}{M_W}\right)^2
\frac{\sqrt{z}( z-4) ^{1.5}}{\left(z-a\right) ^2}\\
\label{dirac_t}
\hat{\sigma}_t^{t}&=\frac{g^4}{32\pi}
\left(\frac{m_\nu}{M_W}\frac{m_{t}}{M_W}\right)^2
\int^{y_{f}}_0 \frac{zy(zy+4)}{(zy+a)^2}dy
\end{align}
where $z\equiv s/m_t^2$ and $a\equiv m_H^2/m_t^2$ are introduced, and the 
limit of integration $y_{f}=(1-m^2_{t}/s)^2$. 

Reduced cross sections of  the diagrams with the $W$ boson in the initial and 
final states are given by 
\begin{align}
\label{dirac_gauge_s}
\hat{\sigma}_s^W&=\frac{g^4}{16\pi}\left(\frac{m_\nu}{M_W}\right)^2
\frac{\left(3-z+z^2/4\right)\sqrt{z(z-4)}}{(z-b)^2}\\
\label{dirac_gauge_t}
\hat{\sigma}_t^W&=\frac{g^4}{16\pi}\left(\frac{m_\nu}{M_W}\right)^2
\int^{y_{g}}_0 \frac{\left(3+zy+z^2y^2/4\right)zydy}{(zy+b)^2}
\end{align}
where $z\equiv s/M_W^2$ and $b\equiv m^2_H/M_W^2$ are introduced, and the limit 
of integration $y_g=(1-M_W^2/s)^2$. 
In the case of  processes with the $Z$ boson $g$ and $M_W$ are to be replaced 
by $\bar{g}$ and $M_Z$ respectively. In addition, overall coefficient of  
expression for the  $s$--channel reduced cross section $\hat{\sigma}_s^Z$ is 
factor of two smaller than that in  (\ref{dirac_gauge_s}).

At the temperature range of interest $T\geqslant m_{H,W,Z,q,\ell}$ and 
approximate analytical expressions for the reaction densities can be obtained. 
To leading order
\begin{align}
\label{dirac_density}
\gamma_s^t&\approx\gamma_t^t\approx \frac{g^4T^4}{512\pi^5}\left(\frac{m_\nu}{M_W}
\frac{m_{t}}{M_W}\right)^2\\
\label{dirac_gauge_density}
\gamma_s^W & \approx 2\gamma_t^W\approx 2\gamma_s^Z\approx 2\gamma_t^Z 
\approx \frac{g^4T^4}{128\pi^5}\left(\frac{m_\nu}{M_W}\frac{T}{M_W}\right)^2
\end{align}
where the  tree--level
relation $gM_Z=\bar{g}M_W$ was used.
\begin{figure}[h]
\begin{center}
\includegraphics{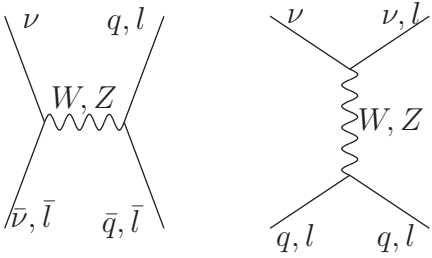}
\end{center}
\caption{\label{diracgaugewashout}Gauge--mediated processes which equilibrate 
lepton asymmetry stored in the left-- and right--handed neutrinos. Here $l$ 
stands for both charged leptons and neutrinos.}
\end{figure}

Diagrams with exchange of the gauge bosons depicted in Fig. \ref{diracgaugewashout} 
also contribute to equilibration of 
lepton asymmetry stored in the left-- and right--handed neutrinos. As 
right--handed neutrino scattering mediated by a gauge boson is a helicity 
flipping process, its amplitude is proportional to the neutrino mass.
Unless the second lepton in the same vertex is also right--handed (amplitude 
of such a process is further suppressed by charged lepton or neutrino mass), 
helicity of a particle in the second vertex is flipped as well, so that the 
amplitude is proportional to mass of this particle. Therefore, the leading 
contribution at the temperature range of interest is due to heavy quarks, in 
particular the $t$--quark. In the case of charged current
\begin{eqnarray}
\label{gaugemed_s}
\hat{\sigma}^{cc}_s&\approx&\frac{g^4}{32\pi}\left(\frac{m_\nu}{M_W}
\frac{m_t}{M_W}\right)^2
\frac{(z-c)^2}{(z-1)^2}\\
\label{gaugemed_t}
\hat{\sigma}^{cc}_t&\approx&\frac{g^4}{32\pi}\left(\frac{m_\nu}{M_W}
\frac{m_t}{M_W}\right)^2
\int_0^{y_f}\frac{zy(zy+c)}{(zy+1)^2}
\end{eqnarray}
where $c\equiv m^2_t/M^2_W$  and $y_f\equiv(1-m_t^2/s)$ are introduced. 
In the case of  neutral current $g$ and $M_W$ are to be replaced by $\bar{g}$ 
and $M_Z$ respectively. In addition, overall coefficient of the  expressions 
for reduced cross sections is factor of two smaller than that in  
(\ref{gaugemed_s}) and (\ref{gaugemed_t}),  $a=0$  and the limit of integration 
$y_f=(1-m_t^2/s)^2$. At the temperature range of interest the corresponding 
reaction densities are given to leading order by
\begin{eqnarray}
\label{gaugemedrd}
\gamma^{cc}_t\approx \gamma^{cc}_s\approx 2 \gamma^{nc}_s \approx 2 
\gamma^{nc}_s \approx\frac{g^4 T^4}{512\pi^5}\left(\frac{m_\nu}{M_W}
\frac{m_t}{M_W}\right)^2 
\end{eqnarray}

Right--hand side of the Boltzmann equations in the expanding Universe 
\cite{Kolb:1990vq} is proportional to the $x\gamma/sH$ ratio. As sphalerons 
are active only at temperatures $T\gtrsim M_W$, it is convenient to define 
$x=M_W/T$, so that $x$ is of order of unity in the relevant range of 
temperatures. Then the ratios are given by
\begin{align}
\label{diracratio}
\frac{x\gamma^t_s}{sH}&\approx \frac{x\gamma^t_t}{sH}\approx
\frac{x\gamma^{cc}_s}{sH}\approx
\frac{x\gamma^{cc}_t}{sH}\approx
2\frac{x\gamma^{nc}_s}{sH}\approx 2\frac{x\gamma^{nc}_t}{sH}
\sim 10^{-13}\\
\hspace*{-2mm}
\frac{x\gamma^W_s}{sH}&\approx 2\frac{x\gamma^W_t}{sH}
\approx 2\frac{x\gamma^Z_s}{sH}\approx 2\frac{x\gamma^Z_t}{sH}\sim 10^{-13}
\left(\frac{T}{M_W}\right)^2
\end{align}
where for the numerical evaluation we used $m_\nu\sim 1$ eV. The reaction 
densities are clearly extremely small and therefore the lepton number asymmetry 
does not come to equilibrium and total number of the right--handed neutrinos 
carrying positive lepton number remain (almost) constant in time.

Neutrino mass eigenstates are related to the 
weak interaction flavor eigenstates by a nondiagonal unitary transformation. 
Provided that  oscillations induced by the neutrino mixing are in thermal 
equilibrium, chemical potentials of different neutrino species are equal. 
Analysis of the neutrino oscillations in the early universe has been performed 
in \cite{Savage:1990by, Dolgov:2002ab} where it was concluded that 
the oscillations begun at about $T\simeq 30$ MeV and flavor equilibrium 
was achieved before the big--bang nucleosynthesis  (BBN) epoch.
At temperatures just below the critical one the oscillations are effectively 
frozen, and the neutrino chemical potentials $\mu_i$ are in general 
not equal, so that  $\Delta_\mu$ differs from zero. 
Conservation of $B-L$ for each family separately (together 
with the rapid flavor--mixing interactions in the quark sector, which assure 
even distribution of baryon number among generations) allows to express
$\Delta_\mu$ in terms of individual lepton numbers $L_i$ 
\footnote{Note that although the individual lepton numbers $L_i$ are not 
conserved, the difference $L_i-L/N$ is conserved by the sphalerons.}.
From (\ref{Bgeneral}) and (\ref{Lgeneral}) it follows that below the 
electroweak phase transition
\begin{align}
(1+2\,c_{\ell_i})\mu_i&=(4-2\Delta_u-2\Delta_d)\,\mu_{uL}\\
&+(2-2c_{i\ell}-2\Delta_d)\,\mu_W-(B/N-L_i)\nonumber
\end{align}
Expanding $\mu_i$ and $c_{\ell_i}\mu_i$ in powers of $1-c_{\ell_i}$ and keeping 
only terms linear in $\Delta_{\ell,u,d}$ we obtain to leading order 
\begin{align}
\label{deltamu}
3\Delta_\mu&=\left[4N\mu_{uL}-(B-L)\right]\Delta_\ell-\Delta\,,\\
\Delta\hphantom{\mu}&=\sum_i\left(L/N-L_i\right)(1-c_{\ell_i})\nonumber
\end{align}
Requiring that  total electric charge is zero and assuming that the 
sphaleron processes are still in equilibrium \cite{Rubakov:1996vz} immediately 
after the phase transition  \footnote{If the phase transition is strongly first 
order, then the anomalous electroweak transitions switch off immediately after 
the phase transition. In this case, obviously, the baryon number is given  by 
its value above the phase transition.} we find:
\begin{subequations}
\label{BandLdirac}
\begin{align}
\label{Bdirac}
B&=\hphantom{-}\frac{8N+4\bar{m}}{24N+13\bar{m}}(B-L)+\Delta B\\
&\approx\hphantom{-	} 0.32\,(B-L)\nonumber+0.59\, \Delta,\\
\label{Ldirac}
L&=-\frac{16N+9\bar{m}}{24N+13\bar{m}}(B-L)+\Delta L\\
&\approx- 0.68\,(B-L)\nonumber+0.59\, \Delta,\\
&\Delta B=\Delta L\equiv\frac{14N+8\bar{m}}{24N+13\bar{m}}\frac{\Delta}3,
\nonumber
\end{align}
\end{subequations}
where  $\bar{m}=(m-1)c_{-}+3c_W$  is introduced.
In comparison with the massless case \cite{Harvey:1990qw} the correction due 
to the Higgs and $W$ mass is about a factor of two. Correction due to the 
fermion masses is about  5\% and neglected  in (\ref{Bdirac}) and (\ref{Ldirac}).
In other words, to obtain relations (\ref{BandLdirac}) we 
neglect $\Delta_{\ell,u,d}$ but keep $\Delta$ and $\Delta_\mu$
different from zero in equations (\ref{BandLgeneral}), 
(\ref{charges}) and (\ref{deltamu}).
For the numerical evaluation we use $N=3$ and assume only one Higgs doublet. 
For definiteness the sphaleron freeze out temperature is chosen to be equal to 
the $W$ mass.

To the same accuracy electric charges of different species are given by
\begin{subequations}
\label{chargesdirac}
\begin{align}
Q_u&=\hphantom{-}\frac{4\bar{m}}{24N+13\bar{m}}(B-L)
+\frac{24N+8\bar{m}}{24N+13\bar{m}}\frac{\Delta}3\\
&\approx\hphantom{-} 0.07\,(B-L)+0.30\,\Delta, \nonumber\\
Q_d&=-\frac{8N+2\bar{m}}{24N+13\bar{m}}(B-L)
-\frac{2N+4\bar{m}}{24N+13\bar{m}}\frac{\Delta}3\\
&\approx -0.28\,(B-L)-0.04\,\Delta, \nonumber\\
Q_l&=\hphantom{-}\frac{8N+6\bar{m}}{24N+13\bar{m}}(B-L)
-\frac{22N+14\bar{m}}{24N+13\bar{m}}\frac{\Delta}3\\
&\approx\hphantom{-} 0.36\,(B-L)-0.32\,\Delta. \nonumber
\end{align}
\end{subequations}
If  $B-L$ is large in comparison  with $\Delta$, then there is an 
excess of down quarks and  total electric charge of the leptons is positive 
immediately after the phase transition.
On the contrary, if $B-L$ is small there is an excess of up quarks and the
total charge of the leptons is negative.

Even if the total $B-L$ asymmetry is zero, resulting baryon ($\Delta B$) and 
lepton ($\Delta L$) numbers are nonzero due to the mass effects.  Let us 
assume for a moment, that the total $B-L$ (more precisely, total initial lepton 
number of the 
left--handed fermions in the scenario under consideration) is zero and the 
final baryon and lepton numbers are nonzero only due to the mass effects. 
Given the smallness of the charged lepton masses, what are the  $L_i$ 
needed to reproduce the observed baryon asymmetry of the Universe?
Taking for the fermion masses their values at zero temperature (which is 
expected to be a good approximation if the phase transition is of the first 
order) we find
\begin{align}
\label{zeroBmL}
B=L&\approx {\cal O} (0.1)\cdot\left[10^{-6}(L/N-L_\tau)\right.\\
&\left.+10^{-8}(L/N-L_\mu)+10^{-13}(L/N-L_e)\right]\nonumber
\end{align}
From (\ref{zeroBmL}) it immediately follows that deviation of individual 
leptons numbers from zero should be at least seven orders of 
magnitude bigger than the observed baryon asymmetry. Large individual lepton 
numbers $|L_i|\gg 10^{-9}$ mean excess of energy density of neutrino sea and 
lead to an increased expansion rate for the Universe which subsequently 
allows less time for the neutrons to decay into protons \cite{Savage:1990by}. 
However,
as temperature drops down to $T_{osc}\simeq 30$ MeV the oscillations equalize 
individual asymmetries (chemical potentials) of the neutrinos. Therefore
chemical potential of the electron neutrino 
at the BBN epoch is determined only by the \emph{total} baryon and lepton 
numbers. 

Let us illustrate this conclusion by estimating the neutrino chemical potential
which is constrained  by BBN and CMBR/LSS  to be in the range 
\cite{Hansen:2001hi}
\begin{equation}
\label{expboundsonxi}
-0.01< \mu_\nu/T < 0.22,
\end{equation}
and the difference $\mu_e-\mu_\nu$ which enters into expression 
for the neutron to proton ratio.  The only leptons at this stage are the 
electron and neutrinos.  Ratio of the electron mass to  temperature 
is still small  and we therefore neglect deviation of $c_e$ from
unity.  Lepton number and electric charge are given by
\begin{equation*}
Y_L=\frac{15}{4\pi^2g^{bbn}_{*}}\left(N\,\frac{\mu_\nu}{T}+2\,\frac{\mu_e}{T}
\right),\,
Y_Q=-\frac{15}{4\pi^2g^{bbn}_{*}}\left(2\,\frac{\mu_e}{T}\right),
\end{equation*}
where $g^{bbn}_*=10.75$ is the effective number of relativistic degrees of 
freedom which are 
in thermal equilibrium with the plasma at the BBN epoch.
Baryons are in form of neutrons and protons at this stage:
\begin{equation}
Y_B=Y_p+Y_n\equiv (1+r)Y_p
\end{equation}
where the neutron to proton ratio $r\approx 1/7$.
Charge neutrality of the Universe implies $Y_Q+Y_p=0$,
and we obtain for the chemical potentials
\begin{align}
\frac{\mu_\nu}{T}&=\frac{4\pi^2g^{bbn}_{*}}{15N}\left( Y_L-Y_p\right)
\hphantom{2)}\\ 
\frac{\mu_e-\mu_\nu}{T}&=\frac{4\pi^2g^{bbn}_{*}}{30N}\left( (N+2) Y_p-
2Y_L\right) 
\end{align}
As follows from (\ref{BandLdirac}),  both $Y_L$ and $Y_B$ are of order of 
$10^{-9}$ and therefore the BBN constrain (\ref{expboundsonxi}) is 
certainly satisfied. The difference $\mu_e-\mu_\nu$ is very small as well,
so that the canonical result for the neutron to proton ratio remains 
unaffected.

Right--handed neutrinos  decouple at a temperature  which 
is well above the temperature of the electroweak phase transition. 
At this stage all the SM degrees of freedom are  
relativistic and new, yet unknown, particle species are very likely to be  
relativistic as well. Standard analysis \cite{Kolb:1990vq} yields an upper 
limit for the 
additional effective  number of neutrino degrees of freedom at the BBN epoch
\begin{equation} 
\Delta N_\nu=N\left( g^{bbn}_{*}/g^{gut}_{*}\right) ^\frac43\leqslant 0.13
\end{equation}  
consistent with the BBN constrain $\Delta N_\nu\leqslant 0.2$ 
\cite{Burles:1999zt}. For numerical evaluation we  used $g^{gut}_*=112$ --- 
effective number of relativistic degrees of freedom in the standard model 
supplemented by three light right--handed neutrinos. 
In  extensions of the SM  $\Delta N_\nu$ is even smaller.

Below the electroweak decoupling  temperatures of
the left-- and right--handed neutrinos scale as 
\begin{equation}
T_{\nu_L}=(g_*/g_*^{dec})^\frac13 T,\quad
T_{\nu_R}=(g_*/g_*^{gut})^\frac13 T
\end{equation}
where the effective number of relativistic degrees of freedom at the 
electroweak decoupling  $g^{dec}_{*}=11/2$. Ratio of the two temperatures
\begin{equation}
T_{\nu_R}/T_{\nu_L}=(g_*^{dec}/g_*^{gut})^\frac13\leqslant 0.36
\end{equation}
The left--handed neutrino become nonrelativistic as temperature drops 
down to $T=(11/4)^\frac13 m_\nu\approx 1.4\, m_\nu$. The right--handed 
one becomes nonrelativistic at a higher temperature $T \simeq 3.9\, m_\nu$.
As the neutrino components are effectively decoupled from the other species, 
their energy density is not transfered to other particles  and therefore 
energy distribution of the nonrelativistic neutrinos deviates from the 
thermal one.

\section{\label{majorana}Leptogenesis with Majorana neutrinos.}
In the scenario of leptogenesis suggested M. Fukugita and T. Yanagida 
\cite{Fukugita:1986hr} CP--violating  decay of heavy 
Majorana neutrino is the source of the initial $B-L$ asymmetry. 
After the EW phase transition the conventional neutrinos 
receive small Majorana masses via the  see--saw mechanism, i.e.
conventional neutrino is a two component Majorana fermion in this
scenario. 

As Majorana fermions do not carry any conserved quantum numbers,
strictly speaking they do not have a chemical potential.  If  this
statement was  applicable in the early Universe, baryon and  
lepton number asymmetries generated above the electroweak 
phase transition would be completely washed out below the 
phase transition \footnote{It is easy to see, that if $\mu_i=0$ below 
the electroweak phase transition, then from the electric charge 
conservation and relation (\ref{sphalerons2}), enforced by the  sphaleron
processes, it follows that $\mu_{uL}=\mu_W=0$ and therefore baryon and 
lepton numbers are zero.}.

For a massless fermion helicity and associated lepton number are
conserved quantities. Presence of the Majorana mass term means 
that there are  processes which flip helicity and violate lepton number 
by two units
\footnote{Note, that above the EW phase transition lepton number 
asymmetry is partially washed out in the scattering processes mediated by 
the heavy Majorana neutrino which violate lepton number by two units.  
However,  these processes are effectively 
frozen out already at temperatures $\sim 0.1 M_N$. Given the current 
estimates of the right--handed neutrino mass $M_N\sim 10^{10}-10^{12}$ GeV 
we conclude that the washout processes are irrelevant in the vicinity of $T_C$.}. 
Rate of such processes is proportional to the neutrino 
mass and, given the current experimental limits \cite{Pakvasa:2003zv, 
Maltoni:2004ei}, is smaller than the expansion rate  of the universe.  
In what follows we assign \cite{Bento:2004xu} positive lepton number to 
neutrinos with left helicity  and negative lepton number to neutrinos with right
helicity  \footnote{For a massive particle helicity is not invariant under 
Lorentz transformations. However, in an isotropic Universe the comoving thermal 
bath frame is a priveleged frame where isotropy enforces the spin density 
matrix to be diagonal in the helicity basis.}.

As has already been mentioned, in the vicinity of $T_C$ only the 
Standard Model states are sufficiently light to be in thermal 
equilibrium. Consequently,  only  coupling of the neutrino to 
the gauge and Higgs bosons are relevant for the present discussion.
\begin{align}
\label{lagrangian}
{\cal L}=&\frac{g}2\frac{m_\nu}{M_W}\,\bar{\nu}\nu^c H_0\\
-&\frac{g}{2\sqrt{2}}\bar{\nu}\gamma^{\mu}\left(1-\gamma^5 \right) e W_\mu
-\frac{\bar{g}}{4}\bar{\nu}\gamma^{\mu}\left(1-\gamma^5 \right) \nu Z_\mu
\nonumber
\end{align}
The first term in (\ref{lagrangian}) is responsible for helicity flipping 
processes, while the second and the third terms contribute to   
two--body scattering mediated by the light neutrino. 

In accordance with the adopted definition of the neutrino lepton number 
helicity flipping processes `directly' violate lepton number by two units and
washout the neutrino chemical potential.
Diagrams with the neutral Higgs  exchange are the same 
(except that now the neutrino is a Majorana fermion) as the ones in 
Fig. \ref{diracwashout}. Expressions for the corresponding reduced cross 
sections and reaction densities are similar to (\ref{dirac_s}) and 
(\ref{dirac_t})  and to (\ref{dirac_density})  respectively. 
Analogously, expressions for reduced cross sections and reaction 
densities  of the diagrams with the gauge boson exchange are similar to 
(\ref{gaugemed_s}) and (\ref{gaugemed_t}) and to (\ref{gaugemedrd}) respectively.

\begin{figure}[h]
\includegraphics{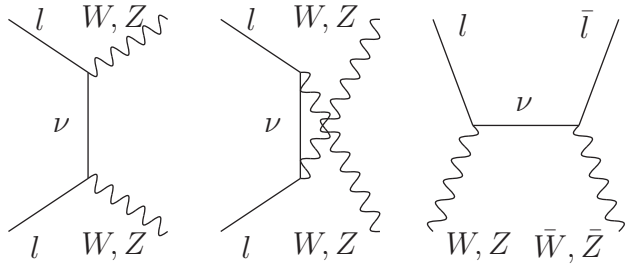}
\caption{\label{majoranawashout}Lepton number violation in two--body processes.}
\end{figure}
There are also several diagrams  with the light neutrino 
exchange which violate lepton number by two units. The neutrino 
chemical potential is washed out `indirectly' by such processes. 
Namely, change of the lepton number induced by these processes
is later transferred to the neutrino sector by the usual SM processes.
Reduced cross sections of the processes (and their charge conjugate) depicted 
in  Fig. \ref{majoranawashout} are given by
\begin{align}
\hat{\sigma}^W_{s}&=\frac{g^4}{512\pi}\left(\frac{m_\nu}{M_W}\right) ^2 
\frac{(z-1)^4}{3 z^7}\\
&\times \left[3z^4+4z^3+58z^2+4z+3 \right] \nonumber\\
\label{tplusu}
\hat{\sigma}^W_{t+u}&=\frac{g^4}{16\pi}
\left(\frac{m_\nu}{M_W}\right)^2
\left(3-z+z^2/4\right)\\
&\times\left[\sqrt{z(z-4)}+\frac{2}{z-2}\ln\left(\frac{z-2+\sqrt{z(z-4)}}{z-2-
\sqrt{z(z-4)}}\right)\right]\nonumber
\end{align}
where $z\equiv s/M_W^2$.  For $T\geqslant m_{W,Z,\ell}$  one can obtain 
approximate analytical expressions for the corresponding reaction densities. 
To leading order
\begin{align}
\gamma_s^W&=\frac{g^4T^4}{(4\pi)^5}\left(\frac{m_\nu}{M_W}\right) ^2 
\left(\frac{T}{M_W}\right) ^2\\
\label{tplusudensity}
\gamma_{t+u}^W&= \frac{9\,g^4T^4}{\pi^5\,\,}\left(\frac{m_\nu}{M_W}\right) ^2 
\left(\frac{T}{M_W}\right) ^6
\end{align}
For the neutral current processes $M_W$ is to 
be replaced by $M_Z$ and $g$ by $\bar{g}$. In addition, overall coefficients of 
expressions for the $t+u$ channel reduced cross section $\hat{\sigma}^Z_{t+u}$ 
and reaction density $\gamma_{t+u}^Z$ are
factor of two smaller than those in (\ref{tplusu}) and (\ref{tplusudensity}).

For  ratios of reaction densities to expansion rate of the universe we find
\begin{equation}
\hspace*{-1mm}\frac{x\gamma^W_{s}}{sH}\sim 10^{-13}\left(\frac{T}{M_W}\right)^2,\
\frac{x\gamma^W_{t+u}}{sH}\sim 10^{-10}\left(\frac{T}{M_W}\right) ^6
\end{equation}
and similar results for the neutral current.

Below the phase  transition lepton number violating interactions are clearly  
too slow to affect the $B-L$ asymmetry generated at the GUT scale, or, 
equivalently, to set the neutrino chemical potential to zero.  Therefore 
the light Majorana neutrino  can be assigned an effective chemical 
potential, and  the rest of the analysis is  the same as in the case of 
leptogenesis with Dirac neutrino. 

Baryon and lepton numbers are given by equations (\ref{BandLdirac}), while
electric charges of different  species are given by equations (\ref{chargesdirac}). 
As has been shown in \cite{Endoh:2003mz, Fujihara:2005pv}
in the leptogenesis with Majorana neutrino scenario  the possibility that 
individual family lepton numbers are much larger than the total lepton 
number is naturaly realized for certain values of the CP violating phases. 
Therefore, terms proportional to $\Delta$ are likely to be important for 
calculation of the lepton and baryon asymmetries.

\section{\label{summary}Summary.}
In this paper lepton and baryon number asymmetries in two scenarios
of leptogenesis have been considered. 

It has been pointed out that if the sphalerons are in thermal equilibrium, 
then the $3B+L$ sum is zero for left--handed fermions.

Despite the fact that a Majorana fermion does not carry conserved quantum 
numbers and does not have a chemical potential, in the early universe the
rate of the processes which bring  number of neutrinos with different 
chirality to equilibrium is 
much smaller than  expansion rate of the Universe, and therefore the 
neutrino can be  assigned an effective chemical potential.

In the scenario of leptogenesis with Dirac neutrinos the smallness of the 
neutrino mass assures that lepton number asymmetries stored in 
the left-- and right--handed neutrinos do not equilibrate.  The right--handed 
neutrinos carry only a small fraction of the Universe energy density and  
have very little impact on the effective number of relativistic degrees of 
freedom, constrained by the BBN. 

Thus the two scenarios are similar from the cosmological 
point of view and yield the same relation among the initial lepton and 
the final baryon asymmetries.

Above the  phase transition baryon and lepton numbers are proportional
to the \emph{total} initial $B-L$ asymmetry. Number of the down quarks exceeds
number of the up quarks and total electric charge of the leptons is positive. 

Below the phase transition due to the mass effects the final baryon number is 
nonzero  even if the initial $B-L$ asymmetry is zero, provided that the 
individual lepton numbers $L_i$ are sufficiently large. As the oscillations 
equalize chemical potentials of the neutrinos long before the BBN epoch, this 
possibility is consistent with the BBN constraints on the effective number of 
relativistic degrees of freedom.

If  $B-L$ is large in comparison  with $\Delta$ then there is an 
excess of down quarks and total electric charge of the leptons is positive 
immediately after the phase transition.
On the contrary, if total $B-L$ is zero, there is an excess of up quarks and
total charge of the leptons is negative.

\begin{acknowledgments}
I would like to thank Professor  E A Paschos for inspiration and useful 
discussions. 
Financial support from the Graduiertenkolleg 841 ``Physik der Elementarteilchen 
an Beschleunigern und im Universum'' at University of Dortmund
is gratefully acknowledged.
\end{acknowledgments}

\end{document}